\documentclass[preprint,aps,nofootinbib]{revtex4}
 \pdfoutput=1

\usepackage{graphicx}

\usepackage{amsmath}
\usepackage{amsfonts}
\usepackage{amssymb}
\usepackage{color}

\def\be{\begin{equation}}
\def\ee{\end{equation}}
\def\bea{\begin{eqnarray}}
\def\eea{\end{eqnarray}}

\def\slashchar#1{\setbox0=\hbox{$#1$}           
   \dimen0=\wd0                                 
   \setbox1=\hbox{/} \dimen1=\wd1               
   \ifdim\dimen0>\dimen1                        
      \rlap{\hbox to \dimen0{\hfil/\hfil}}      
      #1                                        
   \else                                        
      \rlap{\hbox to \dimen1{\hfil$#1$\hfil}}   
      /                                         
   \fi}

\begin{document}
\preprint{EFI-13-14, ANL-HEP-PR-13-34}

\vspace*{1cm}

\title{{ Precision Higgs Measurements: Constraints from New Oblique Corrections}}

\author{\vspace{0.5cm} Stefania Gori$^{\,a,b}$ and Ian Low$^{\,b,c,d}$}
\affiliation{\vspace{0.3cm}
\mbox{ $^a$ Enrico Fermi Institute, University of Chicago, Chicago, IL 60637}\\
\mbox{ $^b$ High Energy Physics Division, Argonne National Laboratory, Argonne, IL 60439}\\
\mbox{ $^c$ Kavli Institute for Theoretical Physics, University of California, Santa Barbara, CA 93106}\\
\mbox{ $^d$ Department of Physics and Astronomy, Northwestern University, Evanston, IL 60208} \\
}

\begin{abstract}
\vspace{1cm}

New particles entering into self-energies of the Higgs boson would necessarily modify loop-induced couplings of the Higgs, if the new particle carries standard model gauge quantum numbers. For a 1 TeV new particle, deviations in these "Higgs oblique corrections" are generically of the order of $v^2/({\rm 1\ TeV})^2\sim$ 5\%. We study constraints on masses and couplings of new scalars and fermions that can be derived from 5--10\% deviations in the Higgs digluon and diphoton partial widths. To reduce theoretical uncertainties, we present next-to-leading order QCD corrections to the Higgs-to-digluon coupling for scalars and fermions in arbitrary representations of $SU(3)_c$ color group, by applying the low-energy Higgs theorems at two-loop order. As a by-product  we provide a new value for NLO QCD corrections to the top squark contributions to digluon decays that differs from existing literature. We also  emphasize that precise measurements of Higgs couplings to $W$ boson and top quark are prerequisite to precise determinations of Higgs oblique corrections from new particles.

\end{abstract}

\maketitle

\section{Introduction}
\label{sect:1}

The discovery of the 125 GeV Higgs boson at the Large Hadron Collider (LHC) at CERN \cite{Aad:2012tfa,Chatrchyan:2012ufa} ushered in the era of "precision Higgs measurements" to determine  properties of the Higgs boson accurately and search for deviations from the standard model (SM) predictions. Such an approach often yields insights into the existence of additional particles not present at  low energies. A prime example is the precision electroweak measurements performed at the LEP at CERN and Tevatron at Fermilab in the past two decades, from which an impressive array of information and constraints were delivered \cite{ALEPH:2005ab}. Among these results  the most striking one  is perhaps predictions of top and Higgs masses using data at the $Z$-pole and measurements of $W$ mass and width: $m_t = 179^{+12}_{-9}$ GeV and $m_h =146^{+241}_{-80}$ GeV. 
Now   the Higgs boson has been discovered,  a program to pursue precision Higgs measurements is  in order.

In addition to "predicting" top and Higgs masses, constraints on possible new physics beyond the SM were also derived from precision electroweak measurements \cite{Altarelli:1990zd}. For a large class of new physics models, corrections to precision electroweak observables are universal \cite{Strumia:1999jm}, in the sense that they only show up in self-energies of electroweak vector bosons. There are strong constraints from these "oblique corrections," pushing the scale of new physics at or above 1 TeV \cite{Barbieri:1999tm}, unless a new parity is imposed \cite{Cheng:2003ju}. Similarly in precision Higgs measurements, corrections to self-energies of the Higgs boson constitute a new class of oblique corrections that are especially sensitive to new physics. They are the focus of this work.

Why do we expect new physics to enter into  the Higgs oblique corrections? The expectation is  based on a theoretical prejudice, ableit a well-founded one, the so-called Naturalness Principle which predicts the existence of new particles to soften the quadratic sensitivity to ultraviolet physics in the Higgs self-energies. One crucial difference from the electroweak oblique corrections, though, is that the Higgs mass at tree-level is a free parameter that cannot be calculated. Therefore  a precise measurement of the Higgs mass will not be able to reveal the size of the Higgs oblique corrections from new physics.

There are, however, other ways these  corrections to Higgs self-energies  would manifest themselves, especially when the new particles carry  SM color and/or electroweak quantum numbers. In these scenarios, for every diagrammatic contribution to the self-energies, one could replace one of the Higgs bosons  by its vacuum expectation value (VEV) and attach two SM gauge bosons to the loop, from which one readily obtains a corresponding diagrammatic contribution to Higgs decays to SM gauge bosons \cite{Low:2009di}. This correspondence is demonstrated in Fig. 1(a) and Fig. 1(b) for new particles carrying electroweak quantum numbers. We see that there is a one-to-one correspondence between Higgs oblique corrections and decay amplitudes for $h\to \gamma\gamma$ and $h\to Z\gamma$. If we  replace the remaining Higgs boson by its VEV again, we see now there is a one-to-one correspondence between the decay amplitudes and the electroweak oblique corrections, as shown in Fig. 1(b) and Fig. 1(c). This correspondence lies in the heart of the low energy Higgs theorems \cite{Ellis:1975ap,Shifman:1979eb}, which relate amplitudes for the loop-induced Higgs decays into two photons and two gluons to the QED and QCD beta functions. Therefore, Naturalness Principle not only predicts new particles entering into the Higgs oblique corrections, but also corrections to the loop-induced Higgs decays. In addition, whether the quadratic sensitivity in the Higgs mass is cancelled or not would dictate the interference pattern between SM and new particles in the loop-induced decays \cite{Low:2009di}. In the end, we see that loop-induced decays of the Higgs boson are the new oblique corrections in precision Higgs measurements.

\begin{figure}[t]
\includegraphics[scale=0.45, angle=0]{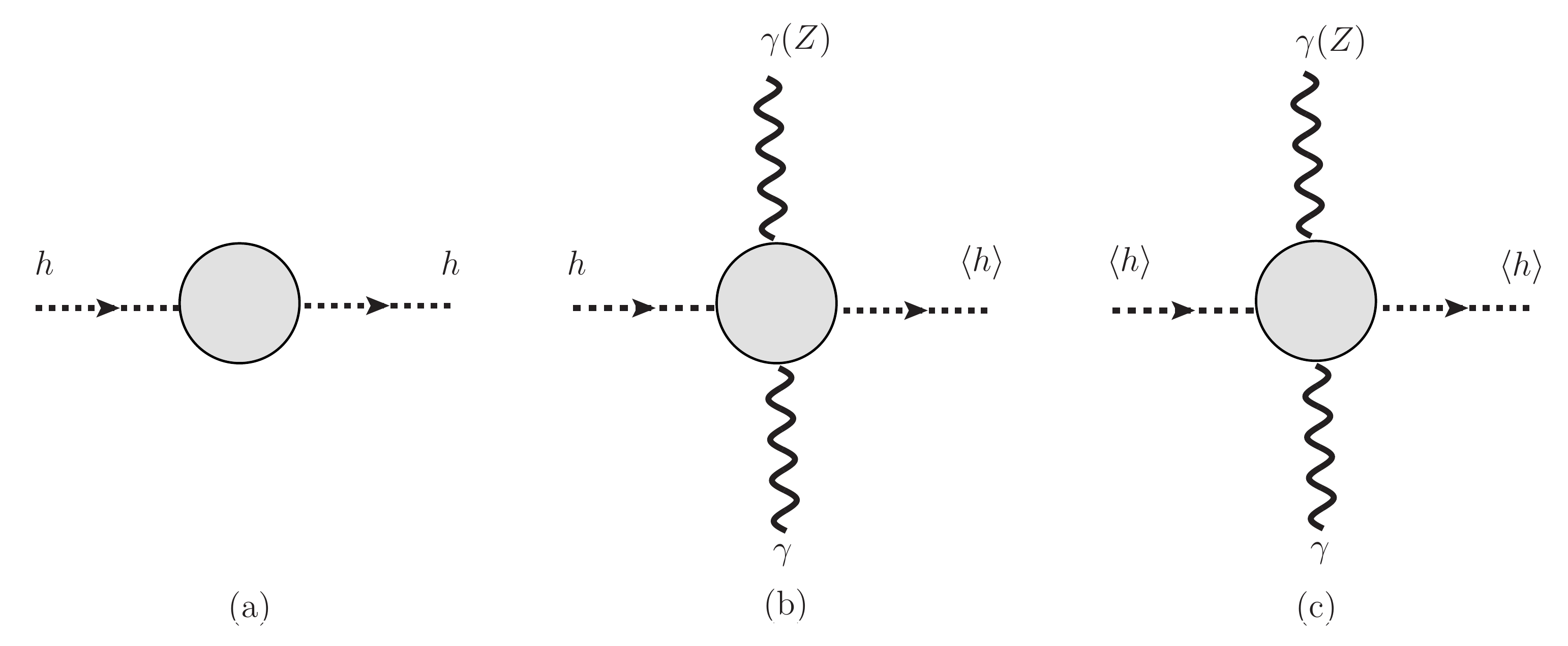}  
\caption{\label{fig1}{\em {\rm (a)} Higgs oblique corrections from new electroweak particles. {\rm (b)} Contributions from new electroweak particles to $h\to \gamma\gamma$ and $h\to Z\gamma$. {\rm (c)} Electroweak oblique corrections from new particles.
}}
\end{figure}

There are many works studying the interplay between new physics and Higgs coupling measurements \cite{Batell:2011pz}. In this work we focus on constraints on masses and couplings of new scalars and fermions from Higgs oblique corrections, concentrating on $h\gamma\gamma$ and $hgg$ couplings.\footnote{$hZ\gamma$ coupling is another Higgs oblique observable that could be measured \cite{Gainer:2011aa}. However the corrections there are generically smaller than in $h\gamma\gamma$ and $hgg$ couplings \cite{Carena:2012xa}.} Before setting out for the actual analysis, it is instructive to estimate the size of possible deviations from precision electroweak measurements, which pushed generic scales of new physics to be at around 1 TeV: $m_{NP} \sim 1$ TeV. Moreover, it is well-known that the dimensionality of operators from integrating out new heavy degrees of freedom start at dimension-six, unless one allows for one dimension-five, lepton number violating operator giving rise to Majorana neutrino masses \cite{Weinberg:1979sa}. It follows that the size of Higgs oblique corrections are roughly given by
\be
\label{eq:size}
{\cal O}\left(\frac{v^2}{m^2_{NP}}\right) \sim 5\% \ ,
\ee
where $v=246$ GeV is the Higgs boson VEV. The above estimate  highlights the importance of making precision measurements.

Precision measurements, however, requires precision predictions. Higher-order corrections in Higgs decays to digluon are known to be substantial \cite{Djouadi:2005gi}. Therefore in this study we take into account next-to-leading order (NLO) QCD corrections in the digluon decays of the Higgs, by applying the low-energy Higgs theorems at two-loop order \cite{Djouadi:1991tka, Dawson:1990zj,Kniehl:1995tn}. In particular, we consider scalars and fermions in arbitrary representations of $SU(3)_c$ color group. On the other hand, since QED corrections to diphoton decays are small, it is sufficient to employ LO results. It is worth noting that, along the way, we provide a new computation for the NLO QCD corrections in top squark contributions to digluon decays of the Higgs, which differs from the literature. Our result applies when all other supersymmetric particles such as the gluino are heavy and decouple. In light of null results from recent collider searches, such an assumption seems quite plausible.

It is also worth emphasizing that constraints and information extracted from precision Higgs measurements are complementary to those derived from direct collider searches, for they involve different sets of assumptions. Direct searches always rely on particular assumptions about the decay modes of the particle being searched for, as well as signal acceptance rates.  For example, searches for final states involving missing transverse energy can always be invalidated by having the would-be missing particle decay further into visible particles. On the other hand, it is also possible to reduce dramatically the signal selection efficiency by choosing some sort of "compressed spectra" so that some, if not all, of the final state particles are too soft to be triggered in detectors. Measurements in the Higgs oblique corrections are not based on any of those assumptions. Nevertheless, it is essential that new particles couple to the Higgs boson in order to show up in Higgs diphoton and digluon decays. Furthermore, if there is more than one new particles, their effects might cancel each other in the loop-induced decays.
Historically it is evident that a two-pronged approach of pursuing both precision measurements and direct searches have worked very well and produced tremendous progress in our understanding of Nature.

This work is organized as follows: in Sec. \ref{sect:2} we consider the leading-order (LO) expressions for the Higgs partial widths in the diphoton and digluon channels in the presence of new particles running in the loop. In Sec. \ref{sect:3} we discuss the importance of including higher order QCD corrections in the digluon channel when one is interested in an accuracy of the order of 5\%. In Sec. \ref{sect.4} we present NLO QCD corrections to the Wilson coefficient of the Higgs-to-digluon coupling for new scalars and fermions in arbitrary representations of QCD color, by applying the low-energy Higgs theorem at two-loop order. Then we consider benchmark scenarios in Sec. \ref{sect.5}, including model-independent scenarios as well as the stop and stau cases in supersymmetry. Sec. \ref{sect.6} contains  discussions on subtleties in relating the partial decay widths to the Higgs couplings in the diphoton and digluon channels, where we also consider an explicit example of composite Higgs models where such subtleties arise. Finally we conclude in Sec. \ref{sect.7}.



\section{Leading-order Decay Widths}
\label{sect:2}

The LO analytic expression for the digluon \cite{Wilczek:1977zn,Georgi:1977gs} and diphoton  \cite{Ellis:1975ap, Shifman:1979eb} partial widths in the SM are
\bea
\label{eq:SMhgg}
\Gamma(h\to gg)&=&\frac{G_F \alpha_s^2 m_h^3}{64\sqrt{2}\pi^3}\left| A_{1/2}(\tau_t) \right |^2 \ ,\\
\label{eq:SMhgaga}
\Gamma(h\to \gamma\gamma)&=&\frac{G_F \alpha^2 m_h^3}{128\sqrt{2}\pi^3}\left|A_1(\tau_W)+ N_c Q_t^2  A_{1/2}(\tau_t) \right |^2 \ ,
\eea
where $G_F$ is the Fermi constant, $N_c=3$ is the number of color, $Q_t=+2/3$ is the top quark electric charge in units of $|e|$, and $\tau_i\equiv 4m_i^2/m_h^2$, $i=t, W$. Below the $WW$ threshold, the loop functions  for spin-1 ($W$ boson) and spin-1/2 (top quark)  particles are given by Eqs.~(\ref{eq:loop1}) and (\ref{eq:loop2}) in the Appendix.


More generally, the LO decay widths in the presence of new particles can be written as \cite{Carena:2012xa}:\footnote{For the purpose of completeness we include the effect of new spin-1 particles in the LO expressions, although in what follows we choose to focus on effects of new scalars and fermions.}
\bea
\label{eq:hgg}
\Gamma(h\to gg)&=&\frac{ \alpha_s^2 m_h^3}{128\pi^3}\left|\delta_{R}\, T(V)\,  \frac{g_{hVV}}{m_V^2}  A_1(\tau_V) +\delta_{R}\, T(F)\, \frac{2g_{hf\bar{f}}}{m_f}  A_{1/2}(\tau_f)\right.\nonumber \\
 && \qquad\left. \phantom{\frac{ \alpha_s^2 m_h^3}{128\pi^3}}  + \delta_{R}\, T(S)\, \frac{g_{hSS}}{m_S^2} A_0(\tau_S) \right |^2  \ , \\
\label{eq:hgaga}
\Gamma(h\to \gamma\gamma)&=&\frac{ \alpha^2 m_h^3}{1024\pi^3}\left|\frac{g_{hVV}}{m_V^2} Q_V^2 A_1(\tau_V) + \frac{2g_{hf\bar{f}}}{m_f} N_{c,f} Q_f^2  A_{1/2}(\tau_f) \right.\nonumber\\
&& \qquad \left. \phantom{\frac{ \alpha_s^2 m_h^3}{128\pi^3}} +  N_{c,S} Q_S^2 \frac{g_{hSS}}{m_S^2} A_0(\tau_S) \right |^2 \ .
\eea
In the above the notation $V$, $f$, and $S$ refer to generic spin-1, spin-1/2, and spin-0 particles, respectively. $T(i), i=V,f,S$ is the Dynkin index of the matter representation defined by the following relation on the group generators:
\be
{\rm Tr}(T^a T^b) = T(i)\, \delta^{ab}\ .
\ee
For  $SU(N)$ fundamental representations and adjoint representations $T(i)=1/2$ and $N$, respectively.
In addition, $\delta_R=1/2$ for real matter fields and 1 otherwise. $Q_V$, $Q_S$  and $Q_f$ are the electric charges of the vectors, scalars and fermions in units of $|e|$, while $N_{c,f}$ and $N_{c,S}$ are the number of fermion and scalar colors. The scalar loop function $A_0$ is defined in Eq.~(\ref{eq:loop3}) in the Appendix. In the limit that the particle running in the loop has a mass much heavier than the Higgs, the loop functions approach
\be
\label{eq:limit}
A_1 \rightarrow b_1= -7  \ , \qquad   A_{1/2} \rightarrow b_{1/2}= \frac{4}3  \ , \qquad A_0 \rightarrow b_0= \frac13 \ ,
\ee
which are related to the one-loop beta functions by the low-energy Higgs theorem \cite{Ellis:1975ap, Shifman:1979eb}. It is often convenient to think of the loop-induced decays as mediated by higher dimensional operators, except in the case of the SM $W$-loop in the diphoton decays. For example, the dimension-five operator responsible for digluon decays of the Higgs in the general case can be written as \cite{Low:2009di}
\bea
\label{generalcoupling}
{\cal L}_{hgg}&=&\frac{\alpha_s}{16\pi}\frac{h}{v}\left [ \delta_R\,  b_{1/2}\, T(f)\, \frac {\partial}{\partial\log v}  \log \det \left({\cal M}_{ {f}}^\dagger {\cal M}_{{f}}\right) \right.\nonumber \\
&&\qquad \ \  +\ \delta_R\,  b_{1}\, T(V)\, \frac {\partial}{\partial\log v}  \log \det \left({\cal M}_{ {V}}^\dagger {\cal M}_{{V}} \right)\nonumber \\
&&\left. \qquad \ \ +\ \delta_R\,  b_{0}\, T(S)\, \frac {\partial}{\partial\log v} \log \det \left({\cal M}_{{S}}^\dagger {\cal M}_{{S}}\right)\right ]G_{\mu\nu}^a G^{a\, \mu\nu} \ ,
\eea
where ${\cal M}_i$ is the mass matrix of the matter particle by turning on the Higgs VEV. For individual particles in the mass eigenbasis, the Higgs coupling to matter particles in Eqs.~(\ref{eq:hgg}) and (\ref{eq:hgaga}) are now
\be
\label{eq:generalcoup}
\frac{g_{hVV}}{m_V^2}= \frac{\partial}{\partial v} \log m_V^2(v)\ , \quad  \frac{2g_{hf\bar{f}}}{m_f}=\frac{\partial}{\partial v} \log m_f^2(v)\ , \quad \frac{g_{hSS}}{m_S^2}= \frac{\partial}{\partial v} \log m_S^2(v)\ .
\ee
In the SM, the $W$ boson and top quark masses are given by
\be
m_W^2 = \frac14 g^2 v^2 \ , \qquad m_t = \frac{1}{\sqrt{2}}\lambda_t v \ ,
\ee
which in turn imply $g_{hWW}=g^2 v/2$ and $g_{ht\bar{t}}=\lambda_t/\sqrt{2}$ and
\be
\label{eq:smcoupling}
\frac{g_{hWW}}{m_W^2}=\frac{2g_{ht\bar{t}}}{m_t}  = \frac{2}v \ .
\ee
Therefore in the SM 
\be
{\cal L}_{hgg}^{(SM)} = \frac{\alpha_s}{12\pi}\frac{h}{v}\, G_{\mu\nu}^a G^{a\, \mu\nu} \ .
\ee

\section{The importance of Higher Order Corrections}
\label{sect:3}

It is well appreciated that higher-order QCD corrections in gluonic decays of the Higgs are substantial, increasing the gluon fusion production rate by more than 50\% at NLO order and another 20-30\% at NNLO order \cite{Djouadi:2005gi}. Heroic efforts have gone into computing the Higgs production and decays precisely. In the SM, the typical approach is to take the heavy mass limit and  "integrate out" the top quark to arrive at a five-flavor effective theory:
\be
\label{eq:Leff}
{\cal L}_{eff} =  {\cal L}_{QCD}^{(5)} + c_{g}\frac{\alpha_s}{12\pi}\frac{h}{v}\, G_{\mu\nu}^a G^{a\, \mu\nu} + \cdots \ ,
\ee
where ${\cal L}_{QCD}^{(5)}$ has the same form as the standard model QCD Lagrangian with five light flavors. Terms omitted in Eq.~(\ref{eq:Leff}) are higher dimensional operators suppressed by powers of $m_t$. The fields and coupling constants in the effective theory are not the same as in the full theory; they are constructed order-by-order in perturbation theory so as to reproduce the $S$ matrix elements of the full theory. This procedure is known as "matching" and usually performed at the scale $\mu=m_t$, below which the effective theory is valid.

Therefore, in Eq.~(\ref{eq:Leff}), $\alpha_s$ is the strong coupling constant in the five-flavor effective theory computed at the scale $\mu$: $\alpha_s=\alpha_s(\mu)$. We take as initial condition \cite{Beringer:1900zz}
\be
\alpha_s(m_Z) = 0.118 \ .
\ee
The Wilson coefficient $c_g$ encodes all the physics at scales above $m_t$ and is computed through the matching. In the SM it has been calculated to N$^3$LO orders  \cite{Djouadi:1991tka, Dawson:1990zj, Kramer:1996iq, Chetyrkin:1997iv}. The NNLO result is simple enough to show:
\bea
\label{eq:CtNNLO}
c_{g}& =& 1 +\frac{11}4 \frac{\alpha_s}{\pi} + \left[\frac{2777}{288} - N_f \frac{67}{96} + \left(\frac{19}{16}+\frac{N_f}{3}\right)\log\frac{\mu^2}{m_t^2} \right] \left(\frac{\alpha_s}{\pi}\right)^2  \ , \nonumber \\
&=&1 + 0.09891 + 0.00796 + \cdots\ ,
\eea
where in the second line above we have evaluated the Wilson coefficient order-by-order in $\alpha_s$ by setting $\mu=m_t$ and $N_f=5$. One sees from Eq.~(\ref{eq:CtNNLO}) that $c_g$ has a nice converging perturbative expansion, where the NNLO correction is below percent level.

The above demonstration suggests the unusually large radiative corrections in the gluonic decay (and gluon fusion production) of the Higgs arise from computing higher corrections {\em within} the effective theory in Eq.~(\ref{eq:Leff}).\footnote{In fact, it is possible to pinpoint the exact origin of large corrections within the effective theory \cite{Ahrens:2008qu}.} As such, the large corrections are associated with degrees of freedom at scales below $m_t$ and insensitive to details of ultraviolet physics encoded in the Wilson coefficient $c_g$. Therefore, if the gluonic decay width of the Higgs is modified by the presence of heavy colored particles, the heavy degrees of freedom would only modify the Wilson coefficient $c_g$, leaving the large universal corrections within the effective theory untouched. In other words, we expect that large QCD corrections should cancel in the ratio of the modified width over the SM width.

 In the effective Lagrangian in Eq.~(\ref{eq:Leff}), only the leading operator in the limit of infinite heavy particle mass is retained. In principle, finite  mass effects could be included systematically by taking into account higher dimensional operators suppressed by $m_t$ \cite{Neill:2009tn}. In practice, however, it is customary to use a hybrid approach by multiplying the LO loop form factors in Eq.~(\ref{eq:hgg}) by the NLO Wilson coefficient \cite{Djouadi:2005gi}:
\bea
\label{eq:hggnlo}
\Gamma^{NLO}(h\to gg)&=&\frac{ \alpha_s^2 m_h^3}{128\pi^3}\, \kappa^{NLO}_{soft}\, \left|
\delta_R\,T(f)\, \frac{2g_{hf\bar{f}}}{m_f}  A_{1/2}(\tau_f)\, c_{f}^{NLO}
\right.\nonumber \\ && \qquad\left. \phantom{\frac{ \alpha_s^2 m_h^3}{128\pi^3}}  
+ \delta_R\, T(R)\, \frac{g_{hSS}}{m_S^2} A_0(\tau_S)\, c_{S}^{NLO}\right |^2  ,
\eea
where $\kappa_{soft}^{NLO}$ is the large QCD corrections due to soft gluons in the effective theory \cite{Baikov:2006ch},
\bea
\kappa_{soft}^{NLO} &=& 1+\frac{\alpha_s}{\pi}\left( \frac{73}4 -\frac76 N_f\right) \nonumber \\
  &=& 1+ 0.427 \ .
 \eea
We see explicitly that the QCD corrections in $\kappa_{soft}^{NLO}$ are very large . However, as explained previously, $\kappa_{soft}^{NLO}$ is agnostic about the heavy degrees of freedom that have been integrated out of the effective theory and hence cancels when taking ratios of a modified gluonic width over the SM expectation.  


Terms neglected in Eq.~(\ref{eq:hggnlo}) include finite $m_t$ effects at NLO, NNLO QCD effects as well as electroweak corrections. In what follows we comment on the importance of these omitted contributions in the ratios of gluonic decay widths.
First let us denote a modified gluonic width $\Gamma$ by
\be
\Gamma = \Gamma_{SM} + \epsilon \ ,
\ee
where $\epsilon$ is the contribution from new heavy particles beyond the SM and generically suppressed by the heavy mass.\footnote{The only exception to decoupling is a 4th generation quark, which would modify the gluon fusion production of the Higgs by an order unity factor and strongly disfavored by current data.} The region of parameter space we are interested here is that the new particle is heavy and the deviation in the width is small,
\be
\label{eq:gameps}
{\cal O}\left(\frac{\epsilon}{\Gamma_{SM}}\right) \alt 10 \% \ .
\ee
Furthermore, we write the difference between NLO widths and the all-order widths as
\bea
\Gamma^\infty &=& \Gamma^{NLO}  + \delta_{SM}^{NLO} + \delta^{NLO} \ , \\
\Gamma_{SM}^\infty &=& \Gamma_{SM}^{NLO}  + \delta_{SM}^{NLO} \ ,
\eea 
where $\delta_{SM}^{NLO}$ corresponds SM corrections that are neglected so far and $\delta^{NLO}$ represents NLO effects from new heavy particles and should be suppressed by the heavy mass. Notice that $\delta_{SM}^{NLO}$ is present in $\Gamma^\infty$ as well. There are three contributions to $\delta_{SM}^{NLO}$: 

\begin{itemize}

\item Electroweak corrections $\delta_{SM,1}^{NLO}$ \cite{Aglietti:2004nj}: ${\cal O}({\delta_{SM,1}^{NLO}}/{\Gamma_{SM}^{NLO}}) \sim 3$\%.\footnote{\label{ft:one}The corresponding number in the reference is for comparing with LO result and therefore slightly larger.}
\item NNLO QCD corrections $\delta_{SM,2}^{NLO}$ \cite{Chetyrkin:1997iv}:  ${\cal O}({\delta_{SM,2}^{NLO}}/{\Gamma_{SM}^{NLO}}) \sim 10$\%.\footnote{See footnote \ref{ft:one}.}
\item Finite $m_t$ mass effects at NLO $\delta_{SM,3}^{NLO}$ \cite{Djouadi:1991tka}: ${\cal O}({\delta_{SM,3}^{NLO}}/{\Gamma_{SM}^{NLO}}) \sim 8$\%.

\end{itemize}
Therefore, it seems conservative to assume that
\be
\label{eq:power1}
{\cal O}\left(\frac{\delta_{SM}^{NLO}}{\Gamma_{SM}^{NLO}}\right) \sim 20\%\ .
\ee
On the other hand, $\delta^{NLO}$ represents contributions to the higher order corrections that are due to the heavy new particles. Given our assumption of heavy particles in Eq.~(\ref{eq:gameps}), we anticipate a similar power counting for ${\delta^{NLO}}/{\delta^{NLO}_{SM}}$:
\be
\label{eq:power2}
{\cal O}\left(\frac{\delta^{NLO}}{\delta^{NLO}_{SM}}\right) \alt 10\% \ .
\ee
Indeed, NLO corrections to the gluonic decay width of the lightest CP-even Higgs in the MSSM is found to be within 5\% of the corresponding SM NLO corrections over most of the parameter space \cite{Harlander:2004tp}: $({\delta^{NLO}_{MSSM}-\delta^{NLO}_{SM}})/{\delta^{NLO}_{SM}} \alt 5\%$. Combining Eqs.~(\ref{eq:power1}) and (\ref{eq:power2}) we arrive at
\be
\label{eq:power3}
{\cal O}\left(\frac{\delta^{NLO}}{\Gamma^{NLO}_{SM}}\right) \sim {\cal O}\left(\frac{\delta^{NLO}_{SM}}{\Gamma^{NLO}_{SM}}\right)\times 
{\cal O}\left(\frac{\delta^{NLO}}{\delta^{NLO}_{SM}}\right) \alt 2\% \ .
\ee
Given these estimates, we can now evaluate errors resulting from terms dropped in Eq.~(\ref{eq:hggnlo}):
\bea
\frac{\Gamma^\infty}{\Gamma_{SM}^\infty} &=& \frac{\Gamma^{NLO}}{\Gamma_{SM}^{NLO}}\left(1 + \frac{\delta^{NLO}}{\Gamma_{SM}^{NLO}} -  \frac{\delta_{SM}^{NLO}}{\Gamma_{SM}^{NLO}}\frac{\epsilon^{NLO}}{\Gamma_{SM}^{NLO}} +\cdots \right)  \nonumber \\
&\sim &  \frac{\Gamma^{NLO}}{\Gamma_{SM}^{NLO}} + {\cal O}(2\%) \qquad \qquad \text{for  ${\epsilon^{NLO}}/{\Gamma_{SM}^{NLO}}\sim 10$\%\ .}
\eea
In the end, we see that, under the assumption in Eq.~(\ref{eq:gameps}), taking ratios of NLO approximation adopted in Eq.~(\ref{eq:hggnlo}) is an excellent approximation when one is interested in a precision in the order of $5-10\%$ in the measurements.

As for new particles contributing to Higgs-to-diphoton coupling, we assume that they do not carry any QCD color. Therefore the NLO corrections arise only in QED, which is of the order of $\alpha/\pi\alt 0.2\%$ and can be safely neglected.

\section{Next-to-leading order Wilson coefficients}\label{sect.4}

In this section we compute the NLO Wilson coefficients in QCD for the digluon decay width by applying the low-energy Higgs theorems at two-loop level \cite{Djouadi:1991tka, Dawson:1990zj,Kniehl:1995tn}, which requires knowledge of two-loop beta functions as well as mass anomalous dimensions of the particle running in the loop. The appearance of the mass anomalous dimension is related to the fact that the counter term necessary to absorb the UV divergence in  the Higgs-loop particles vertex is unambiguously determined by the mass renormalization of the loop particle \cite{Adler:1971pm,Braaten:1980yq}. Since the low-energy Higgs theorem holds in the limit the Higgs momentum vanishes, the Higgs vertex is then renormalized at zero-momentum transfer by an amount $\gamma_m$ \cite{Djouadi:2005gi,Kniehl:1995tn}. Adapting the low-energy theorem to the general case, we arrive at
\bea
\label{eq:higgs2}
{\cal L}_{eff} &=&  {\cal L}_{SM}^{(5)}+ \frac{h}{8v} G_{\mu\nu}^a G^{a\, \mu\nu} \frac{\beta_{\alpha_s}}{\alpha_s} \frac{1}{1+\gamma_m}\ \frac{\partial}{\partial \log v}\log m(v)^2 \ ,
\eea
where ${\cal L}_{SM}^{(5)}$ is the SM lagrangian with five active flavors, $\beta_{\alpha_s} = \partial \alpha_s/ \partial \log\mu$  is the two-loop QCD  beta function, and $\gamma_m$ is the mass anomalous dimension.

Using the results of Ref.~\cite{Jones:1981we}, we can write the two-loop QCD beta function $\beta_{\alpha_s}$  for a fermion in the $f$ representation and a  scalar in the $S$ representation:
\bea
\label{eq:betaf}
{\textcolor{black}{\frac{\beta_{\alpha_s}^{(f)}}{\alpha_s}}}&=&\delta_R\, b_{1/2}\, \frac{\alpha_s}{2\pi}\, T(f) \left\{1+ \frac{\alpha_s}{4\pi}\left[5 C_2(G) + 3 C_2(f)\right]\right\}\ , \\
\label{eq:betas}
{\textcolor{black}{\frac{\beta_{\alpha_s}^{(S)}}{\alpha_s}}}&=&\delta_R\, b_{0}\, \frac{\alpha_s}{2\pi}\, T(S) \left\{1+ \frac{\alpha_s}{2\pi}\left[ C_2(G) + 6 C_2(S)\right]\right\}\ , 
\eea                
where the notations are as follows. For $SU(N_c)$ group the Dynkin index $T(R)=1/2$ for the fundamental representations and $N_c$ for the adjoint representation; $C_2(S)=C_2(f)=C_F$ and $C_2(G)=C_A$ are the quadratic Casimir invariants for the fundamental and adjoint representations, respectively, where $C_F=(N_c^2-1)/(2N_c)$ and $C_A=N_c$. Again $\delta_R=1/2$ for real matter fields, $\delta_R=1$ for complex matter fields and the coefficients $b_i$ are defined in Eq.~(\ref{eq:limit}). 

The mass anomalous dimensions can be obtained in a momentum-independent subtraction scheme such as the $\overline{\rm MS}$  from computing the mass renormalization constant $Z_m$, which relates the renormalized mass parameter $m$ to the bare mass $m_0=Z_m \, m$, in Dimensional Regularization in $d=4-2\epsilon$:\footnote{The definition of $\gamma_m$ in Eq.~(\ref{eq:higgs2}) is such that $\gamma_m = -\partial\log m/\partial \log\mu$.}
\bea
Z_m &=& 1 + Z_{m,1}\, \frac{g_s^2}{\epsilon} + \cdots \ , \\
 \gamma_m& =& -2\times g_s^2 Z_{m,1}\ .
\eea
For fermions in the representation $f$, the calculation of $\gamma_m$ can be found in textbooks \cite{Peskin:1995ev, Srednicki:2007qs}:
\be
\label{eq:gammaf}
\gamma_m^{(f)} = \frac32 \frac{\alpha_s}{\pi} C_2(f) \ .
\ee
For scalars the mass anomalous dimension depends on the scalar quartic coupling defined as follows:
\bea
V(\phi) &=& \frac12 m^2 \phi^a \phi^a +\, \frac{g_s^2}{4!} \lambda_{abcd} \ \phi^a \phi^b \phi^c \phi^d \ , \\
V(\phi,\phi^*) &=& m^2 \phi^{a\, *} \phi^a +\, \frac{g_s^2}{4} \lambda_{abcd}\ \phi^a \phi^{b\, *} \phi^c \phi^{d\, *} \ ,
\eea
where the + sign in front of the quartic is chosen because of vacuum stability constraint and we have normalized the quartic to $g_s^2$. In addition, $\lambda_{abcd}$ are totally symmetric in all four indices for the real scalar and symmetric in the (1,3) and (2,4) indices for the complex scalar. Then we computed
\be
\label{eq:gammams}
\gamma_m^{(S)}=\frac34 \frac{\alpha_s}{\pi} C_2(S) -  \delta_R \frac{\alpha_s}{4\pi} \lambda_{4S}\ , \qquad \lambda_{4S}=\sum_{a} \lambda_{iiaa} \ ,
\ee
where the $i$ index in the quartic is not summed over. Also notice that in the literature the scalar mass anomalous dimension is sometimes quoted as $\gamma_{m^2} = -\partial\log m^2/\partial \log\mu=2\times \gamma_{m}$.\footnote{\label{ft:check}If we consider a single charged scalar in QED, Eq.~(\ref{eq:gammams}) is consistent with the anomalous dimension extracted from the renormalization constants computed in, for example, Ref.~\cite{Srednicki:2007qs} after replacing $\alpha_s\to \alpha$ and $C_2(S)\to 1$. In the case of a (real) color-octet scalar, Eq.~(\ref{eq:gammams}) agrees with the anomalous dimension extracted from Ref.~\cite{Boughezal:2010ry}, if we specialize to the specific form of quartic couplings chosen there: $\lambda_{abcd}=2(\delta_{ab}\delta_{cd}+\delta_{ac}\delta_{bd}+\delta_{ad}\delta_{bc})$. For a (complex) color-triplet scalar, Eq.~(\ref{eq:gammams}) agrees with the squark anomalous mass dimension quoted in Ref.~\cite{Muhlleitner:2006wx}, when the squark quartic interactions are turned-off. However, after including the squark self-interactions from the $D$-term contribution, $\lambda_{abcd}= T^A_{ab}T^A_{cd}+T^A_{ad}T^A_{cb}$, Eq.~(\ref{eq:gammams}) gives $2\alpha_s/(3\pi)$, which is smaller than the number cited in Ref.~\cite{Dawson:1996xz} by a factor of 2. }



Combining the above results, we can now write down the NLO Wilson coefficients for fermions and scalars in arbitrary representations of $SU(3)_c$ using Eq.~(\ref{eq:higgs2}).

\section{Benchmarks}
\label{sect.5}

In this section we consider two classes of benchmark scenarios. The first is assuming only a single particle modifying the digluon and diphoton couplings. For the purpose of illustration we assume that the new color particle in the digluon channel does not carry electroweak quantum numbers, while the new charged particle in the diphoton channel does not carry QCD color. Relaxing these assumptions would only make the constraints stronger because of the the multiplicity of new particles in the loop. In the second class of benchmarks we focus on stops and staus in supersymmetry where there is more than one new particle contributing to the Higgs oblique corrections.

For heavy fermions in the fundamental and adjoint representation, the Wilson coefficients at NLO {\textcolor{black}{entering in eq. (\ref{eq:hggnlo}})} are
\bea
\label{eq:cff}
c^{NLO}_{f(\mathbf{3})}&=&1 +\frac{11}4 \frac{\alpha_s}{\pi}  \ ,\\
c^{NLO}_{f(\mathbf{8})}&=& 1 + \frac32\frac{\alpha_s}{\pi} \ .
 \eea
For scalar particles in the fundamental and adjoint representations, the NLO results are:
\bea
c^{NLO}_{S(\mathbf{3})}&=&  1 + \left(\frac{9}2+\frac{ \lambda_{4S}}4\right)  \frac{\alpha_s}{\pi}\ ,\\
c^{NLO}_{S(\mathbf{8})}&=&  1  + \left(\frac{33}{4} + \frac{\lambda_{4S}}{8} \right)\frac{\alpha_s}{\pi}\ .
\eea
For the squark case, the quartic coupling from the $D$ term\footnote{See footnote \ref{ft:check}.} gives $\lambda_{4S} = C_F =4/3$ and the NLO correction would be 
\be
\label{eq:stopcnlo}
c^{NLO}_{{\tilde t}}=1+ \frac{29}6 \frac{\alpha_s}{\pi},
\ee
 instead of  $1+ {25}\alpha_s/{(6\pi)}$ given by Ref.~\cite{Dawson:1996xz,Anastasiou:2006hc,Djouadi:2005gj}.\footnote{Recently we were informed that there is a sign error in Ref.~\cite{Dawson:1996xz}, which, when corrected, would give a number consistent with Eq.~(\ref{eq:stopcnlo}) \cite{spira}. Furthermore, the authors of Ref.~\cite{Anastasiou:2006hc} pointed out to us that they confirmed the $25/6$ number by
 choosing a minus sign for the stop quartic interaction. When that sign is corrected, they would agree with our result.} For the adjoint scalar with a  particular form of quartic couplings computed in  Ref.~\cite{Boughezal:2010ry}, the above expression gives consistent NLO result.
 
\begin{figure}[t] \centering
\includegraphics[width=0.9\textwidth]{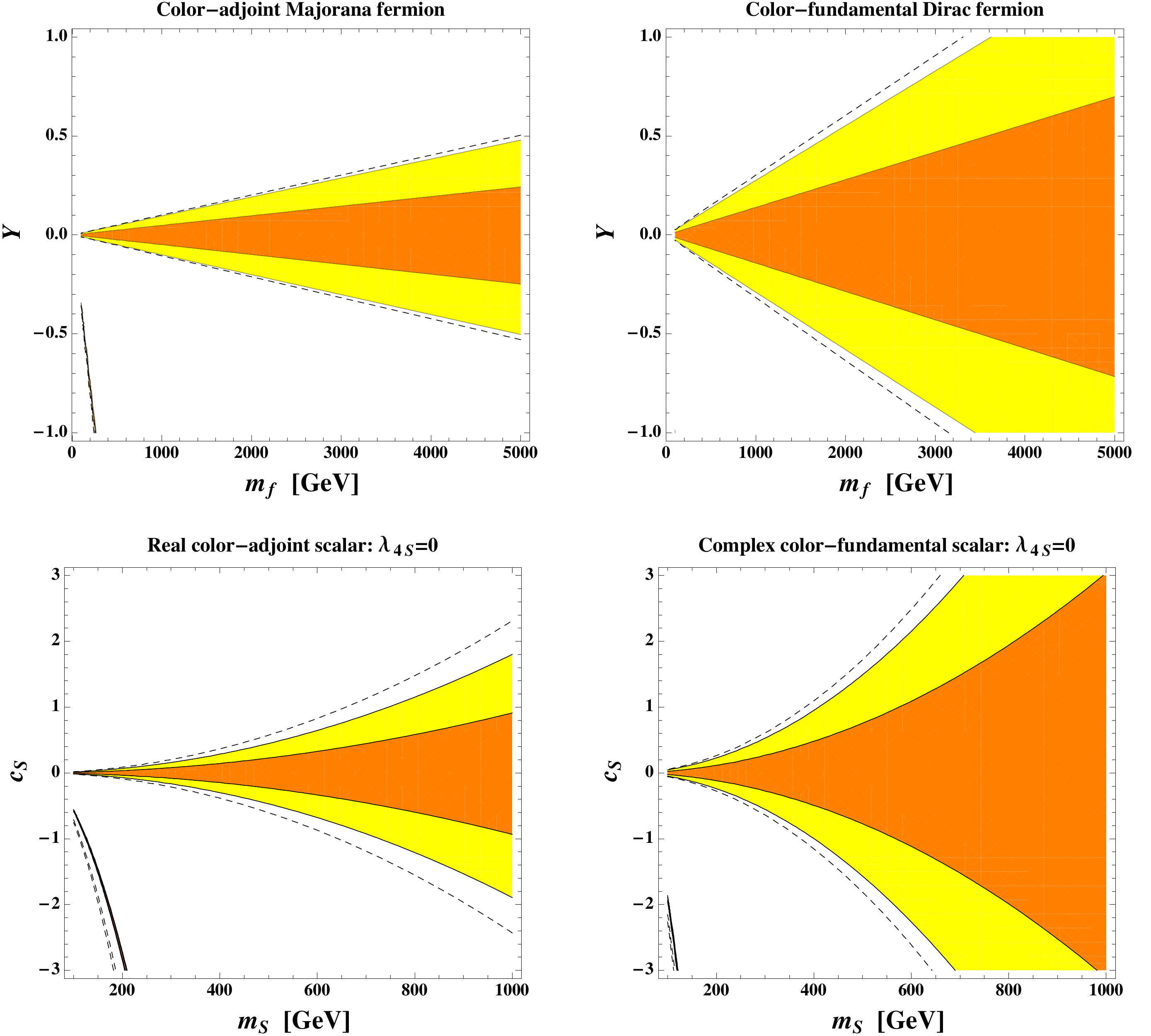}
\caption{\em Contours of constant NLO digluon partial width, normalized to the NLO value in the SM, as a
function of the new particle mass and its coupling to the Higgs. The orange and yellow region are for deviations within  $5\%$ and 10\%, respectively. For comparison, we also show in dashed lines the contour of 10\% deviation from only retaining the LO effect in new particles.}
\label{fig:NPeffectgg}
\end{figure}



In Fig.~\ref{fig:NPeffectgg} we present the constraint on the mass and coupling-to-the-Higgs of a new colored fermion and a new colored scalar from  a $5-10\%$ deviation in $\Gamma(h\to gg)$. In the both  cases we consider a new particle in the fundamental and adjoint representations of $SU(3)_c$ color.  In the fermion case, we assume the coupling of the new fermion to the Higgs  originates from the dimensional-five Higgs-portal operator:
\be
\mathcal O_f=\frac{c_f}{\Lambda}H^\dagger H\bar f f \ .
\ee
After electroweak symmetry breaking, $\mathcal O_f$ generates the following coupling of the new fermion to the Higgs
\be
g_{hf \bar f}=c_f\frac{v}{\Lambda}\equiv\frac{Y}{\sqrt 2}\ .
\ee
For $\Lambda\sim1 $ TeV, typically $Y\lesssim 1$, to guarantee the perturbativity of the $c_f$ coupling until the GUT scale. Measuring the coupling of the Higgs to digluon at the level of $5\%$ would constrain masses up to around 1.2 TeV (1.8 TeV) for fermions in fundamental (adjoint) representation, having fixed $c_f=1$ and $\Lambda=1$ TeV. In the scalar cases, we assume the new scalar  couples to the Higgs through the Higgs portal operator
\be
\mathcal O_S=c_S H^\dagger H S^\dagger S\ ,
\ee
which, after electroweak symmetry breaking, generates the coupling $g_{hSS}=c_S v$. Assuming a coupling $c_S=1$, scalar masses at around 400 (700) GeV in the fundamental  (adjoint) representation of QCD could be excluded by a measurement of the $hgg$ coupling at the level of $5\%$. These numbers have been obtained assuming $\lambda_{4S}=0$. However, we have checked that the reach on the scalar mass does not change significantly for $\lambda_{4S}\sim {\cal O} (1)$.

\begin{figure}[t] \centering
\includegraphics[width=0.9\textwidth]{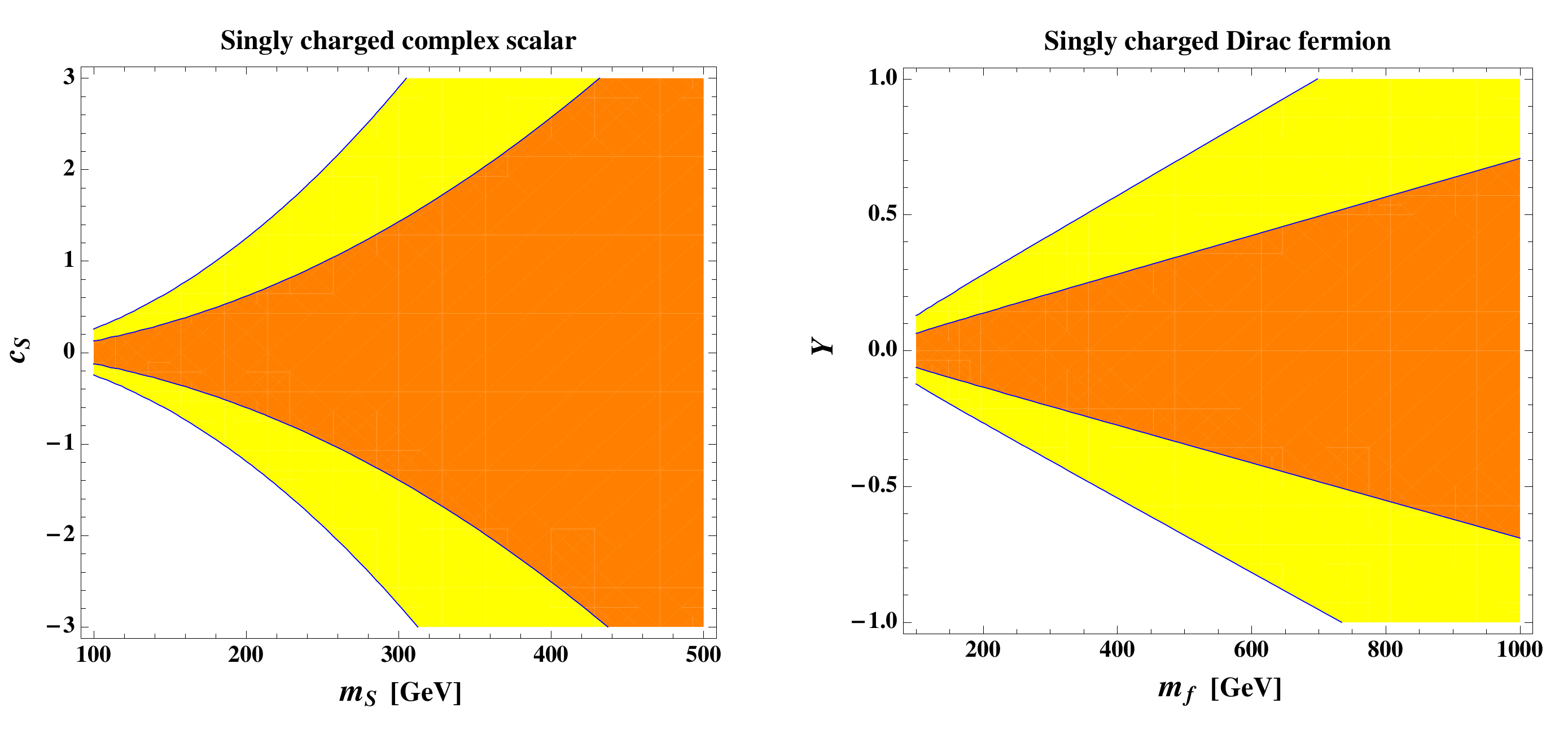}
\caption{\em Contours of constant diphoton partial width, normalized to the SM value. The color-coding is as in Fig.~\ref{fig:NPeffectgg}.}
\label{fig:NPeffectgaga}
\end{figure}

In Fig.~\ref{fig:NPeffectgg} we also show contours of 10\% deviations by taking the ratio of LO partial widths over the NLO SM partial width in the digluon channel, shown as the dashed lines in the plot. The bounds are somewhat weakened in this case. For example, the bound on the mass of the scalar in the adjoint representation would shift from 750 GeV to 650 GeV. It is also important to recall that that the theoretical uncertainties are larger from the arguments presented in Sec. \ref{sect:3}.

In a similar fashion, using the LO results collected in Sec. \ref{sect:2}, we compute the constraint on the $m_f-Y$ and $m_S-c_S$ planes arising from a measurement of the $h\gamma\gamma$ coupling at the $5-10\%$ level. The results are collected in Fig. \ref{fig:NPeffectgaga} .

The constraints on masses and couplings of new particles could change drastically when one relaxes the assumption of only a single new particle in the loop. In supersymmetry there are two top squarks that could modify the digluon width \cite{Djouadi:1998az}, while in the diphoton channel there are two staus that could have important effects \cite{Carena:2011aa}. In the following we consider the interplay of two new particles in the digluon and diphoton partial widths, using stops and staus in supersymmetry as the prime examples.

\begin{figure}[t] \centering
\includegraphics[width=0.9\textwidth]{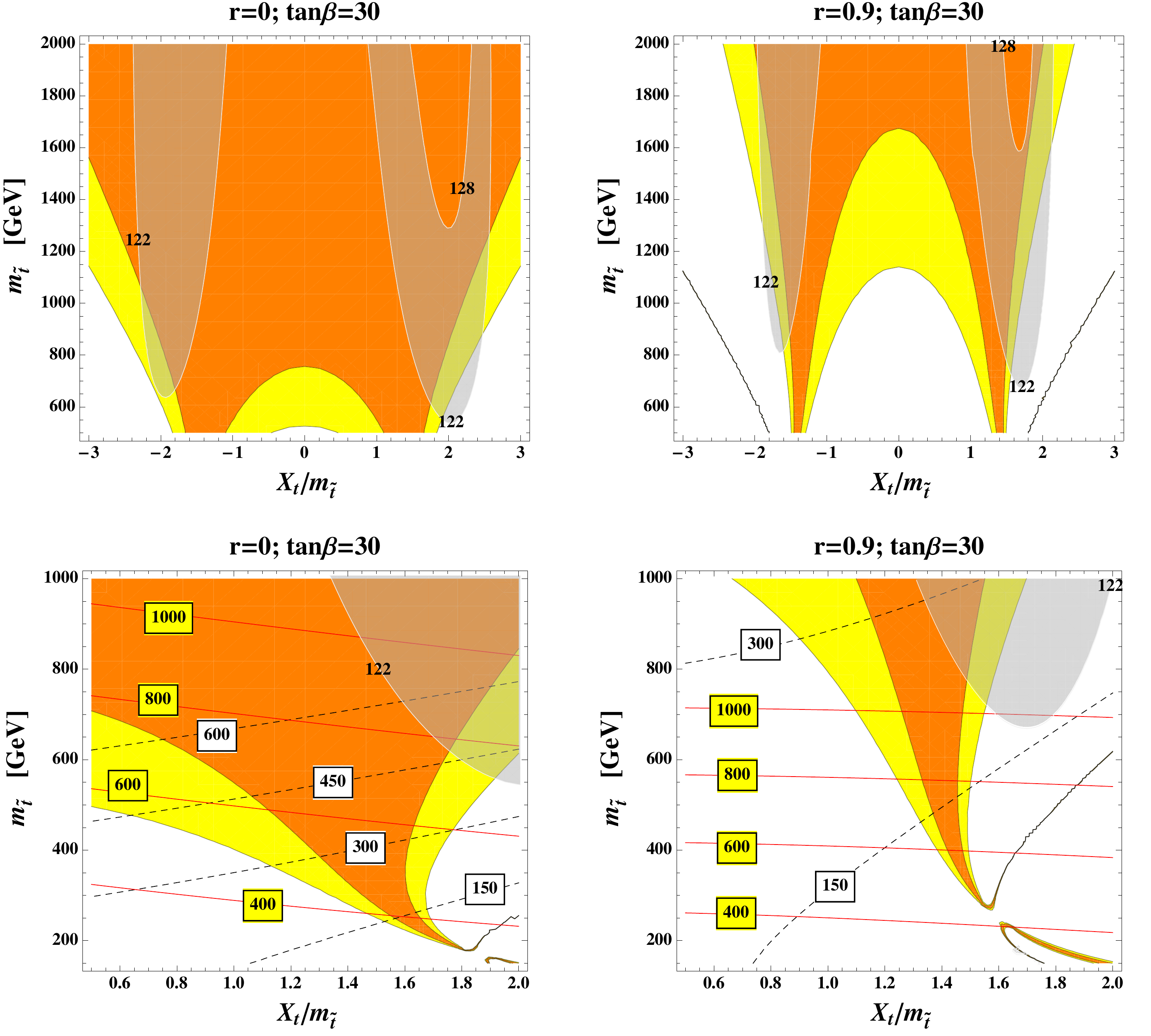}
\caption{\em Contours of constant NLO digluon partial width, normalized to the NLO value in the SM, in supersymmetry as a function of  $m_{\tilde t}$ and $X_t/m_{\tilde t}$. We also include contours of 122 GeV $\le m_h \le$ 128 GeV  in MSSM. In the bottom row we show masses of the light stop eigenmasses (dashed lines) and heavy stop eigenmasses (solid lines). The constraints from digluon widths are independent of MSSM when new $D$ term contributions can be neglected. The color-coding is as in Fig.~\ref{fig:NPeffectgg}.}
\label{fig:NPstops}
\end{figure}

In the stop case, we use the analytic results derived in Sec. \ref{sect.4} by first diagonalizing the stop mass matrix to obtain the mass eigenstates and then including the NLO QCD corrections from Eq.~(\ref{eq:stopcnlo}) for each of the mass eigenstates. The stop mass matrix is written in the flavor basis $(\tilde{t}_L, \tilde{t}_R)$ as:
\begin{equation}
\label{stopmass}
M^2_{\tilde{t}} = \left( \begin{array}{cc} 
   m^2_{\tilde{t}_L} + m_t^2 + D_L^t & m_t X_t \\
    m_t X_t &  m^2_{\tilde{t}_R} + m_t^2 + D_R^t
                         \end{array} \right),
\end{equation}
where
\begin{eqnarray}
D_L^t &=& \left(\frac12-\frac23 s_w^2\right) m_Z^2 \cos 2\beta , \\
D_R^t &=&  \frac23 s_w^2 m_Z^2 \cos 2\beta , \\
X_t &=& A_t - \frac{\mu}{\tan\beta} .
\end{eqnarray}
In the above $s_w$ is the sine of Weinberg angle. We further define
\be
m_{\tilde t}^2 = \frac{m^2_{\tilde{t}_L}+m^2_{\tilde{t}_R}}2 \ , \qquad r = \frac{m^2_{\tilde{t}_L}-m^2_{\tilde{t}_R}}{m^2_{\tilde{t}_L}+m^2_{\tilde{t}_R}}\ .
\ee
We then plot contours of deviations in the digluon width as a function of $m_{\tilde t}$ and $X_t/m_{\tilde t}$, for two extreme values of $r=0$ and $r=0.9$ at $\tan\beta=30$. The outcome is shown in Fig.~\ref{fig:NPstops}.

It should be emphasized that the correlations between $hgg$ couplings and the stop mass matrix in Eq.~(\ref{stopmass}) are  robust predictions of supersymmetry. They are independent of the framework of minimally supersymmetric standard model (MSSM) and would apply to less minimal scenarios such as the Next-to-minimal supersymmetric standard model (NMSSM), when new $D$ term contributions can be neglected. However, for the sake of comparison we also plotted in Fig.~\ref{fig:NPstops} contours of constant Higgs mass in MSSM. Assuming the 125 GeV Higgs is the lightest CP-even Higgs in MSSM and taking into account 3 GeV theoretical uncertainties in computing the Higgs mass in MSSM, we present contours of 122 GeV $\le m_h \le$ 128 GeV computed from {\tt FeynHiggs} \cite{Heinemeyer:1998yj}. It should be noted that we are assuming contributions from other supersymmetric particles such as the sbottom are negligible and much less than 5\% in the digluon width. Under this assumption, we see that, within the framework of MSSM, the constraints derived from digluon widths are sometimes complementary to those obtained from the Higgs mass measurements. In particular, in the region of large $r$, light stop masses would be disfavored if the digluon width is measured to be within 10\% of SM expectation.

\begin{figure}[t] \centering
\includegraphics[width=0.9\textwidth]{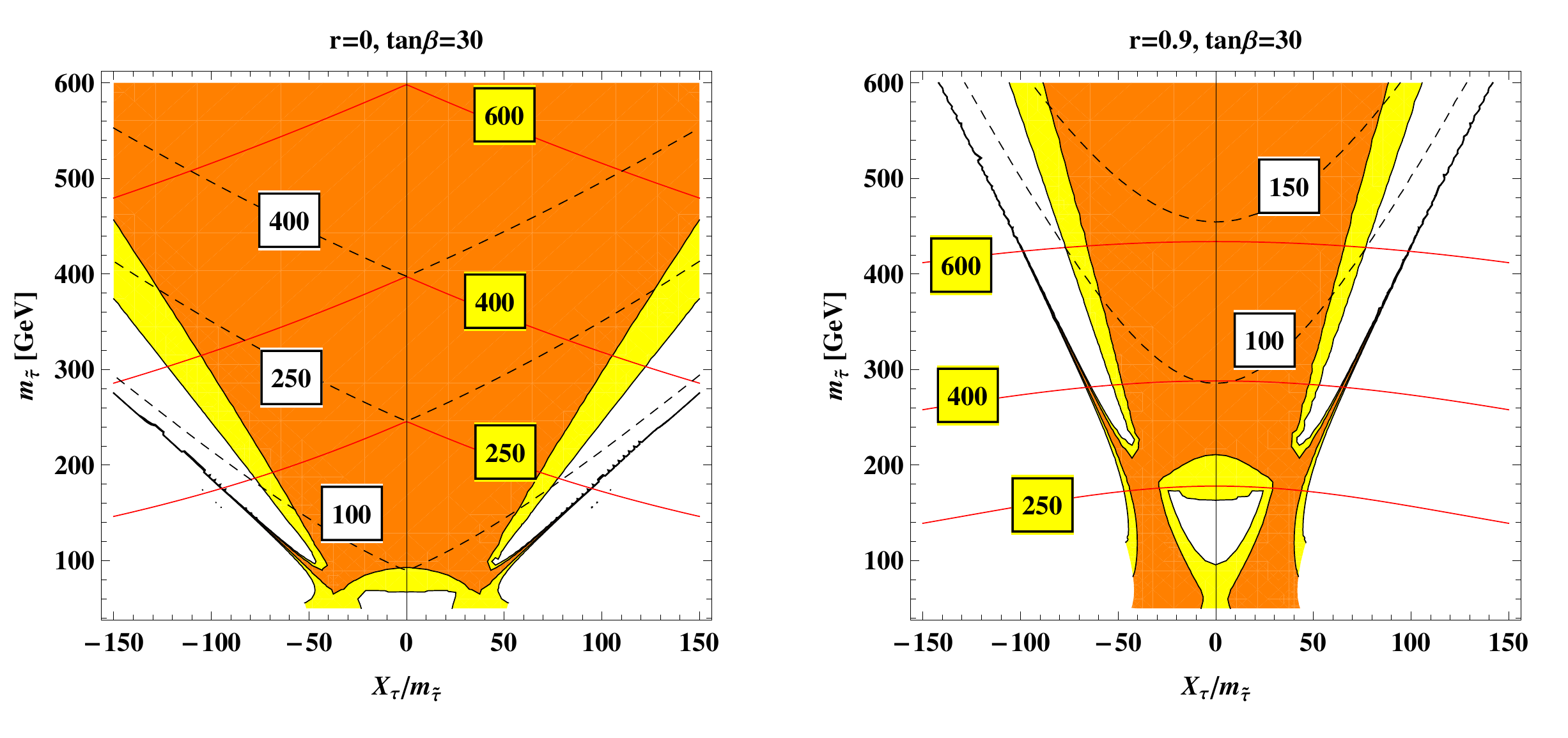}
\caption{\em Contours of constant diphoton partial widths in supersymmetry as a function of $m_{\tilde \tau}$ and $X_\tau/m_{\tilde \tau}$. We also show masses of the light stau eigenmasses (dashed lines) and heavy stau eigenmasses (solid lines). The color-coding is as in Fig.~\ref{fig:NPeffectgg}.}
\label{fig:NPstaus}
\end{figure}

In Fig.~\ref{fig:NPstaus} we show the analogous constraints on the stau sector from Higgs-to-diphoton coupling, where the stau mass matrix is similar to that of the stop in Eq.~(\ref{stopmass}) with the replacement $m_t\to m_\tau$, $m^2_{\tilde{t}_{L/R}}\to m^2_{\tilde{\tau}_{L/R}}$, $X_t\to X_\tau =A_\tau -\mu\tan\beta$, as well as the corresponding changes in the $D$ term contributions. In the plot we assume contributions from other supersymmetric particles such as the charginos can be neglected in the diphoton width.

\section{Subtleties in interpreting partial Widths}
\label{sect.6}

In this section we  comment on subtleties and challenges in extracting diphoton and digluon couplings of the Higgs precisely. 

The main issue is related to the fact that the diphoton and digluon couplings are not  observable experimentally. Instead, what can be measured directly is the diphoton and digluon partial decay widths. In addition to the Higgs oblique corrections discussed so far, the partial widths could be modified if the Higgs couplings to $W$ bosons and top quarks are shifted away from their SM values. These effects could be parametrized as follows \cite{Giudice:2007fh,Contino:2010mh}:
\be
{\cal L} = \frac12 \partial_\mu h\partial^\mu h + 2\frac{m_W^2}{v}\, h W^+ W^-\left(1+ \hat{c}_W\, \frac{v^2}{2\Lambda^2} \right) + 
  \frac{m_t}{v}\, h \bar{t} t \left(1 + \hat{c}_t\,  \frac{v^2}{2\Lambda^2} \right) +\cdots\ ,
  \ee
where $c_W$ and $c_t$ denote deviations of Higgs couplings to $W$ and top from SM values.   
On the other hand, the Higgs oblique corrections that are discussed in this work are summarized by the operators:
\be
\hat{c}_g\, \frac{v }{2\Lambda^2}\, \frac{\alpha_s}{12\pi}\, h G_{\mu\nu}^a G^{\mu\nu\, a} \ , \quad \hat{c}_\gamma\, \frac{v }{2\Lambda^2}\, \frac{\alpha}{6\pi}N_c Q_t^2\, h F_{\mu\nu} F^{\mu\nu} \ ,
\ee
where $N_c=3$ and $Q_t=2/3$ is the top quark electric charge in unit of the electron charge. The the LO digluon and diphoton partials widths are modified at leading order in $v^2/\Lambda^2$ \cite{Giudice:2007fh}:
\bea
\Gamma(h\to gg) &=& \frac{G_F \alpha_s^2 m_h^3}{64\sqrt{2}\pi^3}\left| \left(1+ \hat{c}_t\, \frac{v^2}{2\Lambda^2}\right) \, A_{1/2}(\tau_t) + {\hat{c}_g}\, \frac{v^2}{2\Lambda^2}\, b_{1/2} \right |^2 \ , \\
 &=&\Gamma_{\rm SM} (h\to gg) \left[1 +\left(\hat{c}_t+\frac{\hat{c}_g\, b_{1/2}}{A_{1/2}(\tau_t) }\right) \,  \frac{v^2}{\Lambda^2}\right] +{\cal O}\left(\frac{v^4}{\Lambda^4}\right) \\
 \Gamma(h\to \gamma\gamma)&=&\frac{G_F \alpha^2 m_h^3}{128\sqrt{2}\pi^3}\left|\left(1+ \hat{c}_W\, \frac{v^2}{2\Lambda^2} \right)\, A_1(\tau_W)+ N_c Q_t^2 \left(1+ \hat{c}_t\, \frac{v^2}{2\Lambda^2}\right)  \, A_{1/2}(\tau_t) \right. \nonumber \\
 && \qquad \left. +N_c Q_t^2 \, \hat{c}_\gamma\, \frac{v^2}{2\Lambda^2} \, b_{1/2}  \right |^2 \\
 &=&\Gamma_{\rm SM} (h\to \gamma\gamma) \left[1 +\left( \frac{\hat{c}_W A_1(\tau_W)}{A_1(\tau_W)+N_c Q_t^2 A_{1/2}(\tau_t)} + \frac{\hat{c}_t A_{1/2}(\tau_t)}{A_1(\tau_W)+N_c Q_t^2 A_{1/2}(\tau_t)} \right.\right. \nonumber \\
&&+\left.\left.  \frac{N_c Q_t^2 \, \hat{c}_\gamma\,b_{1/2}}{A_1(\tau_W)+N_c Q_t^2 A_{1/2}(\tau_t)} \right) \frac{v^2}{\Lambda^2}\right]+{\cal O}\left(\frac{v^4}{\Lambda^4}\right) \ .
\eea
We see that, in order to extract the loop-induced couplings $\hat{c}_g$ and $\hat{c}_\gamma$, we need to disentangle their effects in the partial decay widths from those of $\hat{c}_W$ and $\hat{c}_t$.

In terms of $SU(3)_c\times S(2)_L\times U(1)_Y$ gauge invariant operators, $\hat{c}_W$ and $\hat{c}_t$ receive contributions from more than one sources. For example, there is one  operator that contributes to both $\hat{c}_W$ and $\hat{c}_t$ \cite{Giudice:2007fh},
\be
\frac{c_H}{2\Lambda^2} \partial_\mu (H^\dagger H)\partial^\mu (H^\dagger H) \ ,
\ee
while the following two operators contribute to them separately \cite{Giudice:2007fh},
\be
\frac{c_W}{2\Lambda^2} \left(H^\dagger \sigma^i \overset{\leftrightarrow}{D^\nu} H \right) \left(D^\nu W_{\mu\nu}\right)^i \ ,\quad
\frac{c_y}{2\Lambda^2} H^\dagger H (\bar{f} H f)\ .
\ee
There is also an operator $c_{HW}$ that is subleading in weakly-coupled theories  \cite{Giudice:2007fh}.

The operator proportional to $c_H$ is of particular interest, because it gives a finite wave function renormalization to the Higgs boson after electroweak symmetry breaking and one need to re-scale the Higgs field to bring the kinetic term back to canonical normalization:
\be
h \to \frac{h}{\sqrt{1+ c_H v^2/\Lambda^2}} \approx h \left(1-\frac{c_H}2 \frac{v^2}{\Lambda^2}\right) \ ,
\ee
which has the effect of re-scaling all the Higgs partial widths by a universal amount. Moreover, in most cases $c_H>0$ \cite{Low:2009di}, implying all Higgs partial widths receive a universal reduction, while decay branching fractions would remain the same because the Higgs total width is reduced by a similar amount. It was pointed out in Ref.~\cite{Giudice:2007fh} that $c_H$ could be directly measured from longitudinal scatterings of $W$ and $Z$ bosons. For example,
\be
{\cal A}(W^\pm_L W^\pm_L \to W^\pm_L W^\pm_L) = -\frac{c_H\, s}{\Lambda^2}\ ,
\ee
where $s$ is the center-of-mass energy. It would be interesting to consider how $c_H$ could be extracted from Higgs coupling measurements. 

\begin{figure}[t]
\includegraphics[scale=0.9, angle=0]{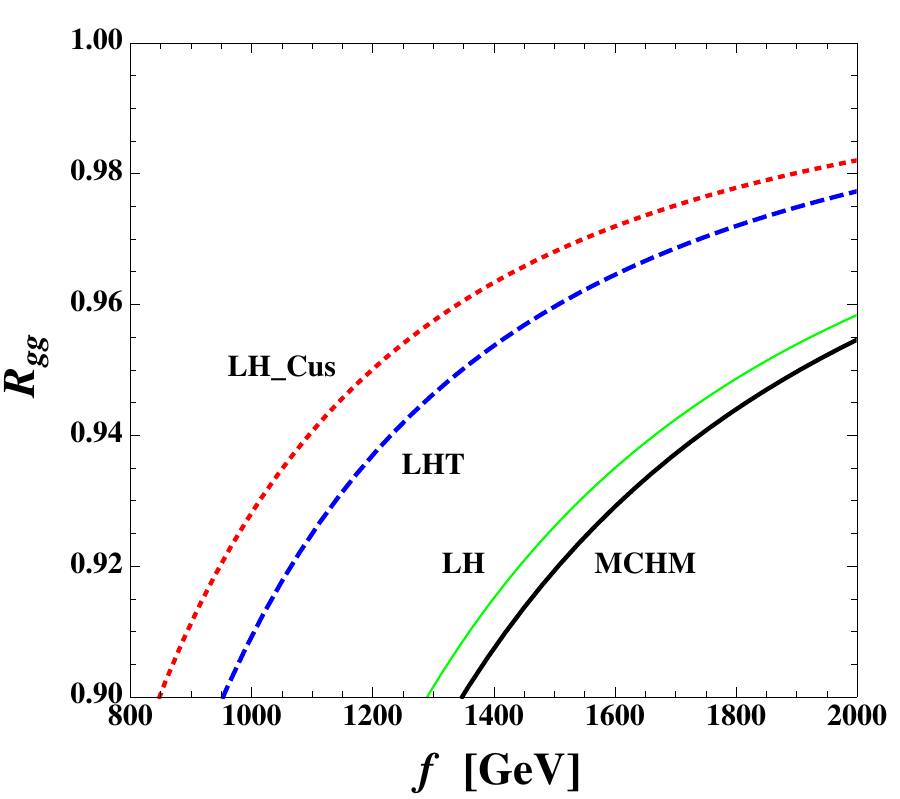}  
\caption{\label{figPNGB}{\em $R_{gg}$ for Higgs as a PNGB. Various curves in the plot represent: the littlest Higgs (LH) model based on $SU(5)/SO(5)$ \cite{Arkani-Hamed:2002qy}, the littlest Higgs with T-parity (LHT) based on $SU(5)/SO(5)\times [SU(2)\times U(1)]^2/SU(2)\times U(1)$ \cite{Pappadopulo:2010jx}, the littlest Higgs with custodial symmetry (LH\_Cus) based on $SO(9)/SO(5)\times SO(4)$ \cite{Chang:2003zn}, and the Minimal Composite Higgs Model (MCHM) based on $SO(5)/SO(4)$ \cite{Agashe:2004rs}. Absence of fine-tunings in the Higgs mass requires $f\alt$ 1 TeV.
}}
\end{figure}

There are in fact very well-motivated models where $\hat{c}_W$ and $\hat{c}_t$ are non-zero. Of particular interest is the possibility that the Higgs boson might arise as a pseudo-Nambu-Goldstone boson (PNGB) \cite{Kaplan:1983fs}, which idea was revived in the past decade \cite{Arkani-Hamed:2001nc,Contino:2003ve}. One rather curious feature of this class of  models is such that, in the Higgs-to-digluon partial widths, effects suppressed by masses of the fermionic top partners disappear due to cancellations in $\hat{c}_g$ and $\hat{c}_t$ \cite{Falkowski:2007hz,Low:2010mr}. In the end the digluon width only depends on $f$, the scale corresponding to the analog of "pion decay constant" in PNGB Higgs. In Fig.~\ref{figPNGB} we show the constraining power of 10\% deviations in $R_{gg}\equiv \Gamma_{\rm PNGB}(h\to gg)/\Gamma_{\rm SM}(h\to gg)$ for a variety of PNGB Higgs models, using the results in Ref.~\cite{Low:2010mr}. It is worth noting that  absence of fine-tunings in the Higgs mass requires $f\alt 1$ TeV.

The observation that $\hat{c}_W$ and $\hat{c}_t$ enters into the Higgs partial decay widths into $gg$ and $\gamma\gamma$ also highlight the importance of making simultaneous precision measurements on Higgs couplings to $W$ bosons and top quarks. While recent studies suggest a precision of the order of 10\% (or less) in $\Gamma(h\to WW)$, $\Gamma(h\to gg)$ and $\Gamma(h\to \gamma\gamma)$ could be achieved either in a high luminosity LHC or an $e^+e^-$ machine, the corresponding precision in the determination of the production rate $\sigma(t\bar{t}h)$ is generally at 10\% or worse \cite{Peskin:2012vu,Klute:2013cx}. Therefore, it becomes apparent that precise measurements of $ht\bar{t}$ coupling should be among the top priorities in the future.

\section{Conclusion}
\label{sect.7}

In this work we argued that Higgs couplings to diphoton and digluon constitute a  class of oblique corrections in Higgs physics, in that new particles entering into self-energies of the Higgs would necessarily induce deviations in Higgs couplings to SM gauge bosons at one-loop order, if the new particles carry SM gauge quantum numbers. Therefore precise measurements of these couplings may yield insights and constraints into masses and couplings of new particles. In particular, for a 1 TeV new particle the resulting deviations are generically of the order of $v^2/({\rm 1\ TeV})^2 \sim$ 5\%, which in turn call for precise theoretical predictions, especially in the digluon couplings where higher-order QCD corrections are known to be substantial.

We then computed NLO QCD corrections from new fermions and scalars in arbitrary representations of $SU(3)_c$ color in the Higgs-to-digluon coupling, by applying the low-energy Higgs theorems. Along the way we present a new computation of the NLO squark contributions. As benchmarks we showed constraints on masses and couplings of new scalars and fermions in the fundamental and adjoint representations of QCD from 5--10\% deviations in the Higgs coupling to digluon. In general constraints on particles in the adjoint representations are more stringent than those in the fundamental representations. In particular,  the allowed region of parameter space for top squarks in supersymmetry was also presented. Similar constraints on charged particles in the Higgs-to-diphoton coupling were presented as well.

Last but not the least, we emphasize that precise determinations of loop-induced couplings of the Higgs require inputs from measurements other than the corresponding partial decay widths. In particular, precise measurements of Higgs couplings to SM $W$ boson and the top quark are necessary to extract diphoton and digluon couplings from the partial decay widths.

In the end, we hope it is clear that much can be learned from a program of precision Higgs measurements, and precise determinations of loop-induced couplings must go hand-in-hand with accurate measurements of other tree-induced couplings. Moreover, indirect probes of new physics from precision Higgs measurements involve different assumptions from direct searches at colliders. Therefore the two approaches are very much complementary to each other and should be pursued simultaneously.


 \begin{acknowledgements} 

We acknowledge helpful discussions and correspondences with Wolfgang Altmannshofer, Thomas Becher, Sally Dawson, Antonio Delgado, Robert Harlander, Bernd Kniehl, Fabio Maltoni, Matthias Neubert and Nausheen Shah. In particular we are grateful to Michael Spira and Babis Anastasiou for clarifying the computations in Ref.~\cite{Dawson:1996xz} and Ref.~\cite{Anastasiou:2006hc}, respectively. This work is supported in part by DOE under Contract No. DE-AC02-06CH11357 (ANL), DE-FGO2-96-ER40956 (U.Chicago), and No. DE-FG02-91ER40684 (Northwestern), and by the Simons Foundation under award No. 230683. Work at KITP is supported by the National Science Foundation under Grant No. PHY11-25915. 
 \end{acknowledgements}

 \section*{Appendix: Definitions of Loop Functions}
 Loop functions used in this paper are defined as follows:
 \bea
 \label{eq:loop1}
A_1(x)&=& -x^2\left[2x^{-2}+3x^{-1}+3(2x^{-1}-1)f(x^{-1})\right]\ .\\
\label{eq:loop2}
A_{1/2}(x) &=& 2  \, x^2 \left[x^{-1}+ (x^{-1}-1)f(x^{-1})\right] \ ,\\
 \label{eq:loop3}
A_0(x) &=& -x^2 \left[x^{-1}-f(x^{-1})\right]  \ , 
 \eea
where
 \be
 f(x) =\left\{
        \begin{array}{cc} 
             \arcsin^2 \sqrt{x} \ , & \ \ x \le 1 \\
             -\frac14 \left( \log\frac{1+\sqrt{1-1/x}}{1-\sqrt{1-1/x}} - i\pi \right)^2\ ,  &\ \  x > 1\ .
             \end{array} 
             \right.
 \ee


\begin{thebibliography}{nn}

\bibitem{Aad:2012tfa} 
  G.~Aad {\it et al.}  [ATLAS Collaboration],
  Phys.\ Lett.\ B {\bf 716}, 1 (2012)
  [arXiv:1207.7214 [hep-ex]].


\bibitem{Chatrchyan:2012ufa} 
  S.~Chatrchyan {\it et al.}  [CMS Collaboration],
  Phys.\ Lett.\ B {\bf 716}, 30 (2012)
  [arXiv:1207.7235 [hep-ex]].
  
\bibitem{ALEPH:2005ab} 
  S.~Schael {\it et al.}  [ALEPH and DELPHI and L3 and OPAL and SLD and LEP Electroweak Working Group and SLD Electroweak Group and SLD Heavy Flavour Group Collaborations],
  Phys.\ Rept.\  {\bf 427}, 257 (2006)
  [hep-ex/0509008];
  [ALEPH and CDF and D0 and DELPHI and L3 and OPAL and SLD and LEP Electroweak Working Group and Tevatron Electroweak Working Group and SLD Electroweak Working Group and Heavy Flavour Group Collaboration],
  arXiv:0811.4682 [hep-ex];
  J.~Alcaraz [ALEPH and CDF and D0 and DELPHI and L3 and OPAL and SLD Collaboration],
  arXiv:0911.2604 [hep-ex];
  [ALEPH and CDF and D0 and DELPHI and L3 and OPAL and SLD and LEP Electroweak Working Group and Tevatron Electroweak Working Group and SLD Electroweak and Heavy Flavour Groups Collaborations],
  arXiv:1012.2367 [hep-ex].


\bibitem{Altarelli:1990zd} 
  D.~C.~Kennedy and P.~Langacker,
  Phys.\ Rev.\ Lett.\  {\bf 65}, 2967 (1990)
  [Erratum-ibid.\  {\bf 66}, 395 (1991)];
  G.~Altarelli and R.~Barbieri,
  Phys.\ Lett.\ B {\bf 253}, 161 (1991);
  P.~Langacker and M.~-x.~Luo,
  Phys.\ Rev.\ D {\bf 44}, 817 (1991);
  M.~Golden and L.~Randall,
  Nucl.\ Phys.\ B {\bf 361}, 3 (1991);
  B.~Grinstein and M.~B.~Wise,
  Phys.\ Lett.\ B {\bf 265}, 326 (1991);
  M.~E.~Peskin and T.~Takeuchi,
  Phys.\ Rev.\ D {\bf 46}, 381 (1992).
 Z.~Han and W.~Skiba,
  Phys.\ Rev.\ D {\bf 71}, 075009 (2005)
  [hep-ph/0412166].

\bibitem{Strumia:1999jm} 
  A.~Strumia,
  Phys.\ Lett.\ B {\bf 466}, 107 (1999)
  [hep-ph/9906266];
  R.~Barbieri, A.~Pomarol, R.~Rattazzi and A.~Strumia,
  Nucl.\ Phys.\ B {\bf 703}, 127 (2004)
  [hep-ph/0405040].
  
\bibitem{Barbieri:1999tm} 
  R.~Barbieri and A.~Strumia,
  Phys.\ Lett.\ B {\bf 462}, 144 (1999)
  [hep-ph/9905281];
   Z.~Han and W.~Skiba,
  Phys.\ Rev.\ D {\bf 71}, 075009 (2005)
  [hep-ph/0412166].
  
\bibitem{Cheng:2003ju} 
  H.~-C.~Cheng and I.~Low,
  JHEP {\bf 0309}, 051 (2003)
  [hep-ph/0308199].
 
\bibitem{Low:2009di} 
  I.~Low, R.~Rattazzi and A.~Vichi,
  JHEP {\bf 1004}, 126 (2010)
  [arXiv:0907.5413 [hep-ph]].

 

\bibitem{Ellis:1975ap} 
  J.~R.~Ellis, M.~K.~Gaillard and D.~V.~Nanopoulos,
  Nucl.\ Phys.\ B {\bf 106}, 292 (1976).


\bibitem{Shifman:1979eb}
  M.~A.~Shifman, A.~I.~Vainshtein, M.~B.~Voloshin and V.~I.~Zakharov,
  Sov.\ J.\ Nucl.\ Phys.\  {\bf 30}, 711 (1979)
  [Yad.\ Fiz.\  {\bf 30}, 1368 (1979)].


\bibitem{Batell:2011pz} 
  N.~Maru and N.~Okada,
  Phys.\ Rev.\ D {\bf 77}, 055010 (2008)
  [arXiv:0711.2589 [hep-ph]].
  F.~Bonnet, M.~B.~Gavela, T.~Ota and W.~Winter,
  Phys.\ Rev.\ D {\bf 85}, 035016 (2012)
  [arXiv:1105.5140 [hep-ph]].
  B.~Batell, S.~Gori and L.~-T.~Wang,
  JHEP {\bf 1206}, 172 (2012)
  [arXiv:1112.5180 [hep-ph]];
  B.~A.~Dobrescu, G.~D.~Kribs and A.~Martin,
  Phys.\ Rev.\ D {\bf 85}, 074031 (2012)
  [arXiv:1112.2208 [hep-ph]];
  A.~Arvanitaki and G.~Villadoro,
  JHEP {\bf 1202}, 144 (2012)
  [arXiv:1112.4835 [hep-ph]];
  D.~Carmi, A.~Falkowski, E.~Kuflik and T.~Volansky,
  JHEP {\bf 1207}, 136 (2012)
  [arXiv:1202.3144 [hep-ph]];
  K.~Kumar, R.~Vega-Morales and F.~Yu,
  Phys.\ Rev.\ D {\bf 86}, 113002 (2012)
  [arXiv:1205.4244 [hep-ph]];
  S.~Dawson and E.~Furlan,
  Phys.\ Rev.\ D {\bf 86}, 015021 (2012)
  [arXiv:1205.4733 [hep-ph]];
  R.~S.~Gupta, H.~Rzehak and J.~D.~Wells,
  Phys.\ Rev.\ D {\bf 86}, 095001 (2012)
  [arXiv:1206.3560 [hep-ph]];
  A.~Joglekar, P.~Schwaller and C.~E.~M.~Wagner,
  JHEP {\bf 1212}, 064 (2012)
  [arXiv:1207.4235 [hep-ph]].
  N.~Arkani-Hamed, K.~Blum, R.~T.~D'Agnolo and J.~Fan,
  JHEP {\bf 1301}, 149 (2013)
  [arXiv:1207.4482 [hep-ph]];
  F.~Bonnet, T.~Ota, M.~Rauch and W.~Winter,
  Phys.\ Rev.\ D {\bf 86}, 093014 (2012)
  [arXiv:1207.4599 [hep-ph]].
  A.~Djouadi,
  arXiv:1208.3436 [hep-ph];
  L.~Wang and X.~-F.~Han,
  Phys.\ Rev.\ D {\bf 87}, 015015 (2013)
  [arXiv:1209.0376 [hep-ph]].
  G.~Passarino,
  Nucl.\ Phys.\ B {\bf 868}, 416 (2013)
  [arXiv:1209.5538 [hep-ph]];
  S.~Dawson, E.~Furlan and I.~Lewis,
  Phys.\ Rev.\ D {\bf 87}, 014007 (2013)
  [arXiv:1210.6663 [hep-ph]];
  T.~Corbett, O.~J.~P.~Eboli, J.~Gonzalez-Fraile and M.~C.~Gonzalez-Garcia,
  Phys.\ Rev.\ D {\bf 87}, 015022 (2013)
  [arXiv:1211.4580 [hep-ph]];
  J.~Reuter and M.~Tonini,
  JHEP {\bf 1302}, 077 (2013)
  [arXiv:1212.5930 [hep-ph]].
  X.~-F.~Han, L.~Wang, J.~M.~Yang and J.~Zhu,
  arXiv:1301.0090 [hep-ph].
  C.~Cheung, S.~D.~McDermott and K.~M.~Zurek,
  JHEP {\bf 1304}, 074 (2013)
  [arXiv:1302.0314 [hep-ph]];
  K.~Cheung, J.~S.~Lee and P.~-Y.~Tseng,
  JHEP {\bf 1305}, 134 (2013)
  [arXiv:1302.3794 [hep-ph]];
  W.~-F.~Chang, W.~-P.~Pan and F.~Xu,
  arXiv:1303.7035 [hep-ph];
  C.~Englert and M.~McCullough,
  arXiv:1303.1526 [hep-ph];
  A.~Joglekar, P.~Schwaller and C.~E.~M.~Wagner,
  JHEP {\bf 1307}, 046 (2013)
  [arXiv:1303.2969 [hep-ph]].
  R.~Contino, M.~Ghezzi, C.~Grojean, M.~Muhlleitner and M.~Spira,
  arXiv:1303.3876 [hep-ph];
  N.~Maru and N.~Okada,
  arXiv:1303.5810 [hep-ph].
  N.~Craig, C.~Englert and M.~McCullough,
  arXiv:1305.5251 [hep-ph].
  X.~-G.~He, Y.~Tang and G.~Valencia,
  arXiv:1305.5420 [hep-ph];
  M.~Farina, M.~Perelstein and N.~R.~-L.~Lorier,
  arXiv:1305.6068 [hep-ph];
  P.~Artoisenet, P.~de Aquino, F.~Demartin, R.~Frederix, S.~Frixione, F.~Maltoni, M.~K.~Mandal and P.~Mathews {\it et al.},
  arXiv:1306.6464 [hep-ph].
  N.~Maru and N.~Okada,
  arXiv:1307.0291 [hep-ph].


\bibitem{Gainer:2011aa} 
  J.~S.~Gainer, W.~-Y.~Keung, I.~Low and P.~Schwaller,
  Phys.\ Rev.\ D {\bf 86}, 033010 (2012)
  [arXiv:1112.1405 [hep-ph]].


  
\bibitem{Carena:2012xa} 
  M.~Carena, I.~Low and C.~E.~M.~Wagner,
  JHEP {\bf 1208}, 060 (2012)
  [arXiv:1206.1082 [hep-ph]].


\bibitem{Weinberg:1979sa} 
  S.~Weinberg,
  Phys.\ Rev.\ Lett.\  {\bf 43}, 1566 (1979).

\bibitem{Djouadi:2005gi} 
 For a summary, see A.~Djouadi,
  Phys.\ Rept.\  {\bf 457}, 1 (2008)
  [hep-ph/0503172].

\bibitem{Djouadi:1991tka} 
  A.~Djouadi, M.~Spira and P.~M.~Zerwas,
  Phys.\ Lett.\ B {\bf 264}, 440 (1991).



\bibitem{Dawson:1990zj} 
  S.~Dawson,
  Nucl.\ Phys.\ B {\bf 359}, 283 (1991);
%
  M.~Spira, A.~Djouadi, D.~Graudenz and P.~M.~Zerwas,
  Nucl.\ Phys.\ B {\bf 453}, 17 (1995)
  [hep-ph/9504378].

\bibitem{Kniehl:1995tn} 
  B.~A.~Kniehl and M.~Spira,
  Z.\ Phys.\ C {\bf 69}, 77 (1995)
  [hep-ph/9505225].

\bibitem{Wilczek:1977zn} 
  F.~Wilczek,
  Phys.\ Rev.\ Lett.\  {\bf 39}, 1304 (1977).
  
\bibitem{Georgi:1977gs} 
  H.~M.~Georgi, S.~L.~Glashow, M.~E.~Machacek and D.~V.~Nanopoulos,
  Phys.\ Rev.\ Lett.\  {\bf 40}, 692 (1978).



\bibitem{Beringer:1900zz} 
  J.~Beringer {\it et al.}  [Particle Data Group Collaboration],
  Phys.\ Rev.\ D {\bf 86}, 010001 (2012).


  

\bibitem{Kramer:1996iq} 
  M.~Kramer, E.~Laenen and M.~Spira,
  Nucl.\ Phys.\ B {\bf 511}, 523 (1998)
  [hep-ph/9611272].


\bibitem{Chetyrkin:1997iv} 
  K.~G.~Chetyrkin, B.~A.~Kniehl and M.~Steinhauser,
  Phys.\ Rev.\ Lett.\  {\bf 79}, 353 (1997)
  [hep-ph/9705240]; 
%
  K.~G.~Chetyrkin, B.~A.~Kniehl and M.~Steinhauser,
  Nucl.\ Phys.\ B {\bf 510}, 61 (1998)
  [hep-ph/9708255].


\bibitem{Ahrens:2008qu} 
  V.~Ahrens, T.~Becher, M.~Neubert and L.~L.~Yang,
  Phys.\ Rev.\ D {\bf 79}, 033013 (2009)
  [arXiv:0808.3008 [hep-ph]].

\bibitem{Neill:2009tn} 
  D.~Neill,
  arXiv:0908.1573 [hep-ph].

\bibitem{Baikov:2006ch} 
  P.~A.~Baikov and K.~G.~Chetyrkin,
  Phys.\ Rev.\ Lett.\  {\bf 97}, 061803 (2006)
  [hep-ph/0604194].

\bibitem{Aglietti:2004nj} 
  U.~Aglietti, R.~Bonciani, G.~Degrassi and A.~Vicini,
  Phys.\ Lett.\ B {\bf 595}, 432 (2004)
  [hep-ph/0404071];
  G.~Degrassi and F.~Maltoni,
  Phys.\ Lett.\ B {\bf 600}, 255 (2004)
  [hep-ph/0407249].

\bibitem{Harlander:2004tp} 
  R.~V.~Harlander and M.~Steinhauser,
  JHEP {\bf 0409}, 066 (2004)
  [hep-ph/0409010].



\bibitem{Adler:1971pm} 
  S.~L.~Adler and W.~A.~Bardeen,
  Phys.\ Rev.\ D {\bf 4}, 3045 (1971)
  [Erratum-ibid.\ D {\bf 6}, 734 (1972)].

\bibitem{Braaten:1980yq} 
  E.~Braaten and J.~P.~Leveille,
  Phys.\ Rev.\ D {\bf 22}, 715 (1980).


\bibitem{Jones:1981we} 
  D.~R.~T.~Jones,
  Phys.\ Rev.\ D {\bf 25}, 581 (1982).


\bibitem{Furlan:2011uq} 
  C.~Anastasiou, R.~Boughezal and E.~Furlan,
  JHEP {\bf 1006}, 101 (2010)
  [arXiv:1003.4677 [hep-ph]]; 
  E.~Furlan,
  JHEP {\bf 1110}, 115 (2011)
  [arXiv:1106.4024 [hep-ph]].


\bibitem{Dawson:1996xz} 
  S.~Dawson, A.~Djouadi and M.~Spira,
  Phys.\ Rev.\ Lett.\  {\bf 77}, 16 (1996)
  [hep-ph/9603423].

\bibitem{Muhlleitner:2006wx} 
  M.~Muhlleitner and M.~Spira,
  Nucl.\ Phys.\ B {\bf 790}, 1 (2008)
  [hep-ph/0612254].

\bibitem{Anastasiou:2006hc} 
  C.~Anastasiou, S.~Beerli, S.~Bucherer, A.~Daleo and Z.~Kunszt,
  JHEP {\bf 0701}, 082 (2007)
  [hep-ph/0611236].



\bibitem{Boughezal:2010ry} 
  R.~Boughezal and F.~Petriello,
  Phys.\ Rev.\ D {\bf 81}, 114033 (2010)
  [arXiv:1003.2046 [hep-ph]].



  
  
\bibitem{Peskin:1995ev} 
  M.~E.~Peskin and D.~V.~Schroeder,
  ``An Introduction to quantum field theory,''
  Reading, USA: Addison-Wesley (1995) 842 p

\bibitem{Srednicki:2007qs} 
  M.~Srednicki,
  ``Quantum field theory,''
  Cambridge, UK: Univ. Pr. (2007) 641 p

\bibitem{Djouadi:2005gj} 
  A.~Djouadi,
  Phys.\ Rept.\  {\bf 459}, 1 (2008)
  [hep-ph/0503173].


\bibitem{spira}
M.~Spira, private communications.

\bibitem{Djouadi:1998az} 
  A.~Djouadi,
  Phys.\ Lett.\ B {\bf 435}, 101 (1998)
  [hep-ph/9806315];
  R.~Dermisek and I.~Low,
  Phys.\ Rev.\ D {\bf 77}, 035012 (2008)
  [hep-ph/0701235 [HEP-PH]].

\bibitem{Carena:2011aa} 
  M.~Carena, S.~Gori, N.~R.~Shah and C.~E.~M.~Wagner,
  JHEP {\bf 1203}, 014 (2012)
  [arXiv:1112.3336 [hep-ph]];
  M.~Carena, S.~Gori, N.~R.~Shah, C.~E.~M.~Wagner and L.~-T.~Wang,
  JHEP {\bf 1207}, 175 (2012)
  [arXiv:1205.5842 [hep-ph]];
  M.~Carena, S.~Gori, I.~Low, N.~R.~Shah and C.~E.~M.~Wagner,
  JHEP {\bf 1302}, 114 (2013)
  [arXiv:1211.6136 [hep-ph]];
  M.~Carena, S.~Gori, N.~R.~Shah, C.~E.~M.~Wagner and L.~-T.~Wang,
  arXiv:1303.4414 [hep-ph].


\bibitem{Heinemeyer:1998yj} 
  S.~Heinemeyer, W.~Hollik and G.~Weiglein,
  Comput.\ Phys.\ Commun.\  {\bf 124}, 76 (2000)
  [hep-ph/9812320];
  T.~Hahn, W.~Hollik, S.~Heinemeyer and G.~Weiglein,
  eConf C {\bf 050318}, 0106 (2005)
  [hep-ph/0507009];
  M.~Frank, T.~Hahn, S.~Heinemeyer, W.~Hollik, H.~Rzehak and G.~Weiglein,
  JHEP {\bf 0702}, 047 (2007)
  [hep-ph/0611326];
  S.~Heinemeyer, W.~Hollik, H.~Rzehak and G.~Weiglein,
  Phys.\ Lett.\ B {\bf 652}, 300 (2007)
  [arXiv:0705.0746 [hep-ph]].



\bibitem{Giudice:2007fh} 
  G.~F.~Giudice, C.~Grojean, A.~Pomarol and R.~Rattazzi,
  JHEP {\bf 0706}, 045 (2007)
  [hep-ph/0703164].

\bibitem{Contino:2010mh} 
  R.~Contino, C.~Grojean, M.~Moretti, F.~Piccinini and R.~Rattazzi,
  JHEP {\bf 1005}, 089 (2010)
  [arXiv:1002.1011 [hep-ph]].

\bibitem{Peskin:2012vu} 
  M.~E.~Peskin,
  arXiv:1208.5152 [hep-ph].


\bibitem{Klute:2013cx} 
  M.~Klute, R.~Lafaye, T.~Plehn, M.~Rauch and D.~Zerwas,
  Europhys.\ Lett.\  {\bf 101}, 51001 (2013)
  [arXiv:1301.1322 [hep-ph]].


\bibitem{Kaplan:1983fs}
D.~B.~Kaplan and H.~Georgi,
Phys.\ Lett.\ B {\bf 136}, 183 (1984);
D.~B.~Kaplan, H.~Georgi and S.~Dimopoulos,
Phys.\ Lett.\ B {\bf 136}, 187 (1984).


\bibitem{Arkani-Hamed:2001nc}
N.~Arkani-Hamed, A.~G.~Cohen and H.~Georgi,
Phys.\ Lett.\ B {\bf 513}, 232 (2001)
[hep-ph/0105239].

\bibitem{Contino:2003ve}
R.~Contino, Y.~Nomura and A.~Pomarol,
Nucl.\ Phys.\ B {\bf 671}, 148 (2003)
[arXiv:hep-ph/0306259].

\bibitem{Falkowski:2007hz}
  A.~Falkowski,
  Phys.\ Rev.\  D {\bf 77}, 055018 (2008)
  [arXiv:0711.0828 [hep-ph]].

\bibitem{Low:2010mr} 
  I.~Low and A.~Vichi,
  Phys.\ Rev.\ D {\bf 84}, 045019 (2011)
  [arXiv:1010.2753 [hep-ph]].

\bibitem{Arkani-Hamed:2002qy}
N.~Arkani-Hamed, A.~G.~Cohen, E.~Katz and A.~E.~Nelson,
JHEP {\bf 0207}, 034 (2002)
[arXiv:hep-ph/0206021].

\bibitem{Pappadopulo:2010jx}
  D.~Pappadopulo and A.~Vichi,
  arXiv:1007.4807 [hep-ph].

\bibitem{Chang:2003zn}
  S.~Chang,
  JHEP {\bf 0312}, 057 (2003)
  [arXiv:hep-ph/0306034].

\bibitem{Agashe:2004rs}
K.~Agashe, R.~Contino and A.~Pomarol,
 Nucl.\ Phys.\ B {\bf 719}, 165 (2005)
[arXiv:hep-ph/0412089].

\end{thebibliography}
\end{document}